\documentclass[preprint,showpacs,preprintnumbers,amsmath,amssymb,floatfix]{revtex4}
\usepackage{graphicx}
\usepackage{dcolumn}
\usepackage{bm}
\usepackage{longtable}
\begin{document}

\title{Origin of transition metal clustering tendencies in GaAs based dilute magnetic semiconductors}
\author{Priya Mahadevan, J.M. Osorio-Guillen and Alex Zunger \\
National Renewable Energy Laboratory, Golden 80401}
\date{\today}

\begin{abstract}
While {\it isovalent} doping of GaAs (e.g. by In) leads to a {\it repulsion}
between the solute atoms, two Cr, Mn, or Fe atoms in GaAs are found to
have lower energy than the well-separated pair, and hence {\it attract}
each other. 
The strong bonding interaction between levels with 
$t_2$ symmetry on the transition metal (TM) atoms
results in these atoms exhibiting a strong tendency to cluster. Using
first-principles calculations, we show that this attraction is maximal for Cr, Mn
and Fe while it is minimal for V. The difference is attributed to the
symmetry of the highest occupied levels. 
While the intention is to find 
possible choices of spintronic materials that show a reduced tendency to cluster, 
one finds that the conditions that minimize clustering tendencies also minimize the
stabilization of the magnetic state.
\end{abstract}


\maketitle

Dilute magnetic semiconductors formed by alloying magnetic 3$d$ ions into
covalent semiconductors have been studied since the eighties 
\cite{az,schneider,clerjaud} and received renewed interest recently
\cite{recent} when high concentration samples ($\sim$ a few percent)
exhibiting ferromagnetism
became available, offering new prospects for spintronic applications. An
important issue here with the high concentration samples 
is the tendency of the magnetic atoms $M$ to
associate \cite{cluster}. To set the background for the problem, let us define the
"substitution energy" $E_{sub}(n)$ as the energy required to take
$n$ atoms of element $M$ from its bulk metallic reservoir (having the chemical potential $\mu_M$)
and use it to replace Ga atoms in GaAs, placing the
ejected Ga atom in its own reservoir (of energy $\mu_{Ga}$):
\begin{eqnarray}
E_{sub}(n)=E[Ga_{N-n}M_nAs_N]-E[Ga_NAs_N]- n \mu_M+ n \mu_{Ga}
\end{eqnarray}
where $E$ is the total energy of the system indicated in 
parentheses, and $N$ denotes the number of atoms. When $E_{sub}(n)$ $>$ 0, substitution
costs energy with respect to solid {\it elemental} sources. For isovalent elements such as $M$=In, 
it was found \cite{cho} that $E_{sub}(1)$ $\sim$ 0.6~eV/cell for 
substitution into bulk GaAs, using the
extreme values of $\mu_{In}$ and $\mu_{Ga}$. For substituting 
Mn in GaAs one similarly
finds E$_{sub}(1)$ $\sim$ 0.9~eV/cell \cite{mn_inst}. 
Thus, substitution costs energy relative to elemental metallic sources.
The substitution energy $E_{sub}(n)$ is related to the 
formation enthalpy $$\Delta H(n)=E[Ga_{N-n}M_nAs_N] -n E[M As] - (N-n) E[GaAs] $$  according to the relation
$E_{sub}(n)=\Delta H(n) + n K$, where,$K=E[M As] - E[GaAs] +\mu_{Ga} -\mu_M. $ 
The calculated $\Delta H$ ($1$)
for dilute Mn in GaAs is 0.37/cell for one  Mn in a 64 atom supercell of GaAs.
Thus, alloying Mn or isovalent In in GaAs
costs energy also with respect to {\it binary} zinc-blende 
(GaAs+MnAs) sources, leading to limited solubility and 
macroscopic phase-separation into GaAs+MnAs at temperatures below the
"miscibility gap" value \cite{epi1}.  
This could be overcome however through surface-enhanced solubility \cite{epi1,epi2} present
during epitaxial growth where the energy of incorporating $M$ at the growing
surface (or near-surface layers) compete favorably with phase separation at the
surface \cite{epi1,epi2}. 

Having introduced In or Mn into the
lattice, one may next inquire whether two such well-separated impurities attract or 
repel each other. For this reason we define the "$M$-$M$ pair interaction
energy" \cite{cho} as the difference in energy of placing two $M$ atoms at
different lattice positions relative to the well-separated limit:
\begin{eqnarray}
\Delta^{(2)}=E[Ga_{N-2}M_2As_N]+E[Ga_NAs_N]
   -2 E[Ga_{N-1}MnAs_N]
\end{eqnarray}
For isovalent alloying of In in GaAs the calculated \cite{cho} repulsion 
was found to be 
$\Delta^{(2)}$
$\sim$ 30~meV/cell for nearest-neighbors along the (110) direction. However,
for two Mn atoms in GaAs an attraction of the order 
$\Delta^{(2)}$ $\sim$ -150 meV has been found in Ref~\cite{mark}. 
Thus, Mn exhibits a thermodynamic 
tendency for atomic association \cite{mark,das}, making the formation 
of "random alloys"
difficult, in contrast with the situation for isovalent semiconductor 
alloys such as GaInAs \cite{cho,epi1}. 
The reason for the tendency of Mn atoms to associate inside a 
III-V semiconductor are however unclear. Schilfgaarde and Mryasov \cite{mark}
concluded that a strong attraction 
arises from the fact that 
the intra-atomic exchange $J$ is large in comparison with
the hopping interaction strength $t$ between the $d$ orbitals. 
Alvarez and Dagotto \cite{dagotto} performed a study of the ferromagnetic
transition temperature $T_c$
as a function of the ratio $J/t$, finding that
for intermediate and large values of this ratio, large ferromagnetic
clusters existed above $T_c$ although long-ranged order was broken. The
basic mechanism responsible for clustering was that when several Mn spins are
close to one another, small regions can be magnetized efficiently. 
These regions remain magnetized even above $T_c$.
Timm and co-workers \cite{timm} suggested that since the 
introduction of Mn in GaAs results in the formation of shallow acceptors, these
generate an attractive Coulomb interaction that favors clustering.

In this paper we inquire as to the physical origin of this 
attraction. We find that all TMs which introduce into GaAs partially
occupied $t_2$ levels leading to ferromagnetism (Cr,Mn), or fully occupied ($t_2$)
levels leading to antiferromagnetism (Fe) inherently tend to cluster ($\Delta^{(n)}$ $<$ 0).
Elements with $e$ levels (V), however, do not introduce strong clustering. Clustering
does not depend on the type of magnetic 
interactions \cite{dagotto}, as it is predicted both for FM and AFM cases.
It also does not depend on acceptors \cite{timm} as it occurs in systems
with deep or shallow acceptors. It is strongest along the $<$110$>$ crystallographic direction.

To evaluate clustering we generalize Eq. (2) to $n$
atoms by calculating 
\begin{eqnarray}
\Delta^{(n)}=[ E(Ga_{N-n}M_nAs_{N}) - E(Ga_{N}As_{N}) ] - n[ E(Ga_{N-1}MAs_{N}) - 
E(Ga_{N}As_{N}) ]. 
\end{eqnarray}
This represents the energy cost for $n$ neutral atoms of type $M$ in a given geometry to form clusters 
relative to the limit in which the atoms are well-separated. In calculating this we use
64 atom supercells of GaAs constructed with 1-4 Ga atoms  
replaced by the transition metal atoms (V/Cr/Mn/Fe). 
Here the lattice constant of the supercell was fixed at the GGA optimized 
value of 5.728 $\AA$ for pure 
GaAs \cite{prb_long}. All atomic positions were relaxed by minimizing 
the total energy as calculated within the plane-wave pseudopotential 
total-energy momentum space method,\cite{ihm_zunger} using ultrasoft pseudopotentials 
\cite{usp}, and the generalized gradient approximation (GGA) \cite{gga} to 
the exchange-correlation as implemented in the VASP code \cite{vasp}. 
We used two types of convergence parameters. 
In the first set (published previously in Ref. \cite{prb_long})
we have used the following convergence parameters: A k-point mesh
of 4x4x4, an energy cut-off of 227.2 eV for Mn, real space projectors, 
no vosko-nusair interpolation scheme and
medium precision in the VASP 
code. This gave 
$\Delta^{(2)}$ of -256, -80, -162 and -206 meV respectively for 1st, 2nd,
3rd and 4th neighbors. These results are plotted in Fig. 1.
In the second set ("highly
converged") we have used a k-point mesh of 4x4x4, an energy 
cut-off of 300 eV, 
Vosko-Wilk-Nusair interpolation scheme for the
gradient term in the exchange functional and accurate
precision in VASP. This gave $\Delta^{(2)}$ of -179, -8, -87 
and -130 meV for 1st, 2nd, 3rd and 4th neighbor Mn. 
The total energies were computed for ferromagnetic as well 
as antiferromagnetic arrangements of the transition metal atoms and 
the lowest energy configuration was 
chosen while evaluating the clustering energy. 
Unless otherwise stated, the calculations have been performed
for the neutral charge state of the defect.

Table I shows our calculated $M$-$M$ pair interaction energies $\Delta^{(2)}$
for nearest neighbor atoms [at (0,0,0) and ($a$/2,$a$/2,0), where $a$ is the
GaAs lattice constant], as well as $\Delta^{(4)}$ for four $M$ atoms located 
at the vertices of the tetrahedron formed by four nearest neighbor
Ga atoms in a zincblende lattice located at (0,0,0), ($a$/2,$a$/2,0), 
($a$/2,0,$a$/2) and (0,$a$/2,$a$/2). We also give in the Table the
electronic configuration of a single $M$ impurity, showing occupation of
$e$-like and $t_2$-like levels \cite{prb_long}. This shows that:

(i) Cr and Mn, having {\it partially occupied} ($t_2$-like) levels
at the Fermi energy as well as Fe with {\it fully occupied} ($t_2$-like)
levels have large attractive pair energies, $\Delta^{(2)}$,
while V having {\it fully occupied} ($e$-type) levels show
significantly reduced tendency to cluster.
Similar tendencies are seen in $\Delta^{(4)}$.
This suggests that the tendency to cluster reflects the nature of the
occupied orbitals on the two impurity atoms. 

(ii) The pair interaction energy $\Delta^{(2)}$ does not correlate with the 
magnetic state, as evidenced by the fact that Cr and Mn pairs are
ferromagnetic while Fe pairs are antiferromagnetic, yet
they both show a strong tendency for clustering. This conclusion contrasts 
with that of Alvarez and Dagotto \cite{dagotto} who associated the clusters
with breakdown of long-range ferromagnetism. By associating 
the formation of clusters with shallow acceptors, Timm \cite{timm} also indirectly 
associated the existence of clusters with the ferromagnetic state, which is not supported by
the present results. 

(iii) The pair interaction $\Delta^{(2)}$ does not correlate with the existence
of shallow acceptor levels, as evidenced by the fact (Table I) that Mn has a 
shallow acceptor in GaAs, but Cr has a deeper one, 
yet $\Delta^{(2)}$ is even more negative for Cr in GaAs. Similarly, the
acceptor in GaN:Mn is extremely deep $E_v$+1.8~eV and $\Delta^{(2)}$ is
found to be extremely negative \cite{mark}.  
This conclusion contrasts with that of Timm \cite{timm}, who suggest that 
long-ranged attractive 
Coulomb interactions produced by uncompensated shallow 
acceptor producing defects bring about the clustering. 
These shallow acceptor producing defects induce an attractive force between the nuclear core
of $M$ and the bound hole.
As the Bohr radius for shallow acceptors is large, the wavefunction
of the hole could overlap with that of another similarly bound hole
about another $M$ present. Hence the energy lowering is greater in the
case when the acceptor level is shallower.

(iv) The pair interation $\Delta^{(2)}$ does not 
correlate with the $J$/$t$ ratio. Indeed, the strength of the 
coupling $t$ of $d$ orbitals with $e$ symmetry
on neighboring TM atoms is weaker than between orbitals with $t_2$ symmetry
because in the zincblende structure, while the $t_2$ orbitals 
point to those on the  neighboring atom, 
the $e$ orbitals point  at an angle of 45$^{\o}$ to the line joining them \cite{az}
As the magnitude of $J$ is not expected to change across the series
V-Fe, the ratio $J$/$t$ is larger for V in GaAs, than it is for Cr-Fe in
GaAs. However, Table I shows that 
the clustering tendencies do not follow the trend of the ratio $J$/$t$.

(v) We have also performed calculations to examine clustering 
tendencies in the charged states of the defects. Recent experiments \cite{goldman}
find a tendency of such defects to anticluster. Considering the case of 
two Mn$_{Ga}^{-1}$ defects that are stable when  
the Fermi energy is above the acceptor level at E$_v$+0.1 eV,  we find that $\Delta^{(2)}$
for nearest neighbor pairs 
is reduced to -70~meV from -256~meV for Mn$_{Ga}^0$ pairs. The reduction
could have two origins. The first being that the 
repulsion between the charged Mn$_{Ga}^-$ units destabilizes the 
formation of clusters. The second is that the antiferromagnetic state
associated with the a pair of Mn$_{Ga}^-$ atoms occupying nearest neighbor 
Ga positions is weakly stabilized ($\sim$ 120~meV/cell). 

{\it What are the energetics favoring clustering?} The strong dependence
of clustering  on the symmetry of the highest occupied orbital 
suggests that the large values of the intraatomic exchange interaction 
strength $J$ in comparison with the bonding strengths $t$ are certainly 
not the origin. The dependence on the symmetry arises because the hopping 
interaction strength $t$ between two transition metal atoms 
are different for $e$ and $t_2$ symmetries. 
The states with $e$ symmetry 
on the TM atom have no counterparts on the host lattice to couple to, so the
TM($e$)-TM($e$) coupling is rather weak. In contrast the states with $t_2$
symmetry on the TM can couple to host states of the same symmetry
available at the same energy range, so strong {\it indirect} TM($t_2$)-host($t_2$)-TM($t_2$)
effective coupling exists.

The presence of clusters of 2-4 Mn atoms are difficult to detect.
Our results suggest that
the tendencies for TM clustering in GaAs is intrinsic. It is difficult to
suppress clustering during growth (as {\it interstitial}
Mn can be suppressed by annealing of a {\it thin} film), as the substitutional 
clusters are not mobile at annealing temperature.

{\it Strong directional dependence of the matrix elements:} The 
coupling between states with $t_2$ symmetry 
will be largest for two TM atoms occupying lattice positions 
along the zincblende bonding chain {\it i.e.} joined by the
translation vector ($a$/2,$a$/2,0), while it would be the smallest 
when the translation vector is ($a$,0,0).
On the other hand, for states with $e$ symmetry, the hopping matrix 
elements would be largest
when the lattice vector joining the atoms is along the ($a$,0,0) 
direction, and smallest along the ($a$/2,$a$/2,0)
direction. Consequently nearest-neighbor Ga-substitutional
positions will not be favored when the highest occupied level has $e$ symmetry.
We make quantitative estimates of this aspect of clustering by 
considering pairs of transition metal atoms with the first atom at 
the origin and the second at ($a$/2,$a$/2,0) $\equiv$ NN1;
or ($a$,0,0) $\equiv$ NN2, or ($a$/2,$a$/2,$a$) $\equiv$ NN3, or ($a$,$a$,0) $\equiv$ NN4
being the NN-th neighbor.
The clustering/pairing energy were evaluated and the results are plotted in Fig.~1. 

We see indeed that : (i)
the results for Cr, Mn and Fe indicate that the strengths of the hopping matrix elements are largest when the atoms
can be joined by the vector along the (1 1 0) direction. (ii)
It is not just nearest neighbor
lattice positions that are mutually attractive, but even farther neighbor
Mn pairs show substantially negative $\Delta^{(2)}$. 
(iii) Clustering is favored by the magnetic ground state whether FM (Cr, Mn) or
AFM (Fe), whereas magnetically {\it excited} states (AFM - Cr , AFM - Mn or FM -Fe) have 
weaker clustering tendencies. This is because 
a substantial portion of the energy favoring 
clustering comes from the energy stabilizing the observed magnetic ground state.
The clustering energy is not equal to the magnetic stabilization energy 
as there is an energy cost brought about by the additional
perturbation of the host lattice in bringing two or more impurity atoms close to each other
compared to when they are far separated. 

We conclude that clustering is produced by the tendency of 
$t_2$ orbitals on each TM to couple, thus lowering the energy
of the system. This tendency is maximal for 
bond-oriented $M$-$M$ pairs. Note that the magnetism 
itself is stabilized by the same
bonding interaction. Thus, systems with weak clustering (eg V) also
have weak magnetism.

We acknowledge support from the Office of Naval Research. We thank Y.J. Zhao
for useful discussions on the subject.

\newpage
\begin{table}
\caption 
{ Clustering energy (Eq. (3))  and the favored magnetic configuration 
for pairs and for 4 atom clusters of transition metal atoms.
Results are given per 64-atom cell. 
The "formal" electronic configuration as well as location of acceptor
transitions for isolated impurities are also provided. The VASP 
convergence parameters correspond to "set 1" defined in the text.} 
\vspace{1cm}
\begin{tabular}{cccccc}
TM & $\Delta^{(2)}$ (in meV) & $\Delta^{(4)}$ (in meV) & FM/AFM & config. & Acceptor \\ \hline
V  & -31               & -31   & FM & $e^2$ & \\ \hline
Cr & -281              & -1086 & FM & $e^2t^1$ & $E_v$ +0.74 \\ \hline
Mn & -256              & -795  & FM & $e^2t^2$ & $E_v$ +0.11 \\ \hline
Fe & -304              & -708  & AFM & $e^2t^3$ &  \\ \hline
\end{tabular}
\end{table}

\newpage 
\clearpage

\begin{figure}
\caption{ The pairing energies (Eq. (2)) for 2 V, Cr, Mn and Fe atoms in GaAs at 1-4 neighbor Ga-substitutional 
positions for FM (black squares) and AFM (black circles) arrangement of their spins. The results have been calculated using "set 1" defined in the text.
}
\end{figure}
\end{document}